\documentclass[12pt]{article}
\usepackage{amsfonts,amssymb,amsmath}
\usepackage[dvips]{epsfig}
\begin{document}
\title{\bf\large{Minimum-error discrimination between two sets of similarity-transformed quantum states}}\vspace{20mm}
\author{M. A. Jafarizadeh, $^{1,2,3,}$
 \thanks{E-mail:jafarizadeh@tabrizu.ac.ir}
 Y. Mazhari Khiavi,$^{1,}$
 \thanks{E-mail:mazhari@tabrizu.ac.ir} Y. Akbari Kourbolagh, $^{4,}$ \thanks{E-mail:yakbari@azaruniv.edu}
\\
$^{1}${\small Department of Theoretical Physics and
Astrophysics, University of Tabriz, Tabriz 51664, Iran.}  \\
$^2${\small Institute for Studies in Theoretical Physics and
Mathematics, Tehran 19395-1795, Iran.}\\$^{3}${\small Research
Institute for Fundamental Sciences, Tabriz 51664, Iran.}\\
$^{4}${\small Department of Physics, Azarbaijan University of Tarbiat Moallem, 53714-161 Tabriz, Iran.}}

\pagebreak
 \newtheorem{thm}{Theorem}
 \newtheorem{cor}[thm]{Corollary}
 \newtheorem{lem}[thm]{Lemma}
 \newtheorem{prop}[thm]{Proposition}
 \newtheorem{defn}[thm]{Definition}
 \newtheorem{rem}[thm]{Remark}
\vspace{20mm}

\maketitle

\begin{abstract}
Using the  new form of necessary and sufficient conditions introduced in Ref. \cite{Jafarizadeh2}, minimum error discrimination among
two sets of similarity transformed equiprobable quantum qudit states is investigated. In the case that the unitary operators are
generating sets of two irreducible representations, the optimal set of measurements and the corresponding maximum
success probability of discrimination are determined in closed form. In the case of reducible representations, there exists no closed-form formula
in general, but the procedure can be applied in each case accordingly. Finally, we give the
maximum success probability of optimal discrimination for some important examples of mixed quantum states,
such as qubit states together with three special cases and generalized Bloch sphere m-qubit states.

{\bf PACs Index: 01.55.+b, 02.10.Yn }
\end{abstract}

\section{Introduction}
The discrimination of nonorthogonal quantum states is an important and challenging problem in the
realm of quantum information theory. It is known that one cannot discriminate perfectly among nonorthogonal
quantum states and ought to invoke to an optimal way of state
discrimination. There exist two approaches to the problem of discrimination: optimal minimum-error discrimination in which the state identification is probabilistic \cite{Helstrom1}-\cite{Assalini1} and optimal unambiguous discrimination in which the states are discriminated without error \cite{Ivanovic1}-\cite{Jafarizadeh1}. Here, we consider discrimination strategies, known as the minimum-error discrimination (MED), which are based
upon the minimization of the rate error. For treating minimum error
discrimination, there exist two known strategies : the necessary and sufficient conditions for optimal discrimination \cite{Helstrom1}-\cite{Barnett2}
and Helstrom family of ensembles \cite{Kimura1}. However, solving problems by means of them, except for some particular cases, is a difficult task. In Ref. \cite{Jafarizadeh2},
a new technique has been presented in which the two previous strategies are comprised to obtain optimality conditions
of the equality form and it has been shown that the new technique is powerful in solving
problems of optimal discrimination between mixed quantum states which are in general not
symmetric. Here, we use the latter technique in the problem of minimum error discrimination among two different sets of
equiprobable similarity transformed quantum states.
\par In this paper, we first use the technique of Ref. \cite{Jafarizadeh2} to
investigate MED between two sets of equiprobable quantum mixed states
generated from two original density operators by unitary similarity transformations.
The novelty of this work is that quantum states to be discriminated are partitioned into two sets with non-equal probabilities. In the case that the unitary
operators are irreducible representations of generators of a subgroup of unitary group
$U(d)$, the maximum
success probability and optimal measurement operators are precisely derived. In the
case that the unitary
operators are reducible representations of the corresponding subgroup, although there exists in general no closed-form formula but
the procedure can be applied in each case accordingly. Finally, we study MED between some
important classes of mixed quantum states such as generalized Bloch sphere m-qubit states,
in details.
\section{Minimum error discrimination among states of two sets of similarity transformed equiprobable states} In general, the
measurement strategy is described in terms of a set of
positive semidefinite operators $\Pi_i$ known as the probability operator
measure (POM). The measurement outcome labeled by $i$ is associated with the operator $\Pi_i$ and the sum of the POM elements must be
the identity operator, i.e., $\sum_i\Pi_i=\mathrm{I}$ which is known as a resolution of the identity. Knowing that the transmitted state is $\rho_j$, the
probability of observing the outcome $i$ by the receiver is $p(i|j)=Tr
(\Pi_i\rho_j)$.
Let us consider the different states
$\rho_{1},\rho_{2},\ldots,\rho_{N}$ with prior probabilities
$p_{1},p_{2},\ldots,p_{N}$, respectively $(p_{i}\geq
0,\sum_{i}p_{i}=1)$. Then the success probability $p$ for correctly
identifying the states $\rho_i$ is given by
\begin{equation}\label{e1}
p=\sum^{N}_{i=1}p_{i}Tr(\rho_{i}\Pi_{i}).
\end{equation}
\par The necessary and sufficient conditions leading to the
minimum-error discrimination can be written as \cite{Jafarizadeh2}
\begin{equation}\label{e2}
\mathcal{M}=p_{j}\rho_{j}+(p_{opt}-p_{j})\tau_{j},\;\;\;\
\;\ j=1,\ldots, N.
\end{equation}
in which $\mathcal{M}$ denotes $\sum^{N}_{i=1}p_{i}\Pi_{i}\rho_{i}$, $p_{opt}$ stands for maximal success probability, $p_{opt}\geq p_{j}$ and $\tau_j$'s are positive operators of trace one called conjugate states \cite{Kimura1}.
If $\Pi_{k}=I$ for some $k\in\{1,\ldots N\}$, then taking trace of both sides of Eq. (\ref{e2}) gives $p_{opt}=p_{k}$.
Note that for two different equiprobable states $\rho_{k}$ and $\rho_{l}$ it is impossible that $p_{opt}=p_{k}=p_{l}$, since in this case Eq. (\ref{e2}) implies $\rho_{k}=\rho_{l}$ which is a contradiction.
It is shown in Ref. \cite{Jafarizadeh2} that the conditions (\ref{e2})
lead to
\begin{equation}\label{e3}
(p_{opt}-p_{j})Tr(\tau_j\Pi_j)=0
,\;\;\;\
\;\ j=1,\ldots, N;
\end{equation}
\par
In this paper, we consider two sets of $N$ different states: the first set containing equiprobable unprimed states with prior probabilities $\eta$ and the second set containing equiprobable primed states with prior probabilities $\eta'$ as follows
\begin{equation}\label{e6}
\rho_{j} =U_j\rho_1U_j^{-1},\;\;\ \tau_{j}=U_j\tau_1U_j^{-1},\;\;\ \;\ j=1,\ldots,n,
\end{equation}
\begin{equation}\label{e7}
\rho'_{j} =U'_{j}\rho'_{1}{U'}^{-1}_{j},\;\;\ \tau'_{j}=U'_{j}\tau'_{1} {U'}^{-1}_{j},\;\;\  \;\ j=1,\ldots,n',
\end{equation}
where $n+n'=N$ and $\{U_1=I_d,U_2,\ldots, U_{n}\}$ and $\{U'_{1}=I_d,U'_{2},\ldots, U'_{n'}\}$ are  generating sets of
representations of two subgroups of $U(d)$, the group of all $d\times d$ unitary matrices. Clearly, we have
\begin{equation}
\quad n\eta+n'\eta'=1.
\end{equation}
Such states is referred to as similarity transformed states. Denoting POM elements corresponding to $\rho_{j}$ and $\rho'_{j}$ by $\Pi_{j}$ and $\Pi'_{j}$ respectively, we must have a resolution of the identity
\begin{equation}\label{e9}
\sum^{n}_{j=1}\Pi_j+\sum^{n'}_{j=1}\Pi'_{j}=I.
\end{equation}
Assume that $\Pi_j=\lambda_{j}\pi_{j}$ and $\Pi'_j=\lambda'_{j}\pi'_{j}$ for some nonnegative numbers $\lambda_{j}$ and $\lambda'_{j}$ and positive
semidefinite operators $\pi_{j}$ and $\pi'_{j}$ such that $\sum^{n}_{j=1}\lambda_{j}+\sum^{n'}_{j=1}\lambda'_{j}=1$. As POM operators constitute a resolution of the identity, it is possible that for the optimal POM some of the operators $\Pi_{i}$ and $\Pi'_{i}$ vanish. Note here that $\pi_{j}$'s and $\pi'_{j}$'s must be satisfy the conditions (\ref{e3})
\begin{equation}\label{e9-3}
(p_{opt}-\eta)\lambda_{j}Tr(\tau_j\pi_j)=(p_{opt}-\eta')\lambda'_{j}Tr(\tau'_j\pi'_j)=0
\end{equation}
so the optimal operators $\pi_{j}$ and $\pi'_{j}$
can also be assumed to obtain from positive operators $\pi_{1}$ and $\pi'_{1}$ respectively
via the same similarity transform which defines the states $\rho_{i}$ and $\rho'_{i}$ and the corresponding conjugate states $\tau_{i}$ and $\tau'_{i}$,
i.e.
\begin{equation}\label{e9-1}
\pi_{i}=U_{i}\pi_{1}U^{-1}_{i},
\end{equation}
\begin{equation}\label{e9-2}
\pi'_{i}=U_{i}\pi'_{1}U^{-1}_{i}.
\end{equation}
Therefore, Eq. (\ref{e9-3}) is reduced to
\begin{equation}\label{e9-3-1}
(p_{opt}-\eta)\lambda_{j}Tr(\tau_1\pi_1)=(p_{opt}-\eta')\lambda'_{j}Tr(\tau'_1\pi'_1)=0.
\end{equation}
Here, optimality conditions (\ref{e2}) can clearly be written as
\begin{equation}\label{e9-4}
\begin{array}{c}
  \mathcal{M}=\eta\rho_{i}+(p_{opt}-\eta)\tau_{i} \\
\hspace{1cm}=\eta'\rho'_{j}+(p_{opt}-\eta')\tau'_{j}.
\end{array}
\end{equation}
where $\mathcal{M}$ denotes $\eta\sum^{n}_{i=1}\Pi_{i}\rho_{i}+\eta'\sum^{n'}_{i=1}\Pi'_{i}\rho'_{i}$.
From (\ref{e6}), (\ref{e7}) and (\ref{e9-4}), it is easy to see that
\begin{equation}\label{e9-5}
\mathcal{M}=U_i\mathcal{M} U_i^{-1}=U'_j\mathcal{M} {U_j'}^{-1}.
\end{equation}
We know that the representations $U_{i}$'s and $U'_{j}$'s are either irreducible or reducible. We discuss the two cases separately.\\\\

\subsection{The irreducible case}
Let $U_i$'s and $U'_j$'s be generating sets of
irreducible representations of two subgroups of $U(d)$. Then, by the Schur's first lemma on representation theory
\cite{Joshi1,James}, Eq. (\ref{e9-5}) implies that $\mathcal{M}$ is a multiple of identity
operator; i.e., $\mathcal{M}=\alpha
I$ for some complex number $\alpha$.\\
By taking trace of Eq. (\ref{e9-4}), we get
\begin{equation}\label{e13}
\alpha=\frac{p_{_{opt}}}{d}.
\end{equation}
In order to obtain an
optimal measurement for discrimination among states $\rho_i$, let us write $\rho_1$ in the spectral decomposition form as
$\rho_1=\sum_{i=1}^da_i| i\rangle\langle i|$.
Then, Eq. (\ref{e9-4})
implies that $\tau_1$ is also diagonal in the basis $|
i\rangle$, say, $\tau_1=\sum_{i=1}^db_i|i\rangle\langle
i|$. By the replacement of $\rho_1$ and $\tau_1$ in Eq. (\ref{e9-4}), we have
\begin{equation}\label{e14}
\frac{p_{opt}}{d}\sum_{i=1}^d|
i\rangle\langle
i|=\eta\sum_{i=1}^da_i|
i\rangle\langle
i|+(p_{opt}-\eta)\sum_{i=1}^db_i|
i\rangle\langle i|,
\end{equation}
and hence
\begin{equation}\label{e15}
p_{opt}=d[\eta a_i+(p_{_{opt}}-\eta)b_i].
\end{equation}
Similarly, for the second set, we have
\begin{equation}\label{e16}
p_{opt}=d[\eta'a'_i+(p_{_{opt}}-\eta')b'_i].
\end{equation}
Since, for the irreducible case we have $n\geq2$ and $n'\geq2$, then as mentioned above $p_{opt}\neq\eta,\eta'$ and hence Eq. (\ref{e9-3}) implies that
\begin{equation}\label{e17}
\lambda_{j}Tr(\tau_{1}\pi_{1})=\lambda'_{j}Tr(\tau'_{1}\pi'_{1})=0.
\end{equation}
Hence, either $\lambda_{j}=0$ ($\lambda'_{j}=0$) or $\pi_{1}$ ($\pi'_{1}$) is  perpendicular to $\tau_{1}$ ($\tau'_{1}$); namely, $\tau_{1}$ ($\tau'_{1}$) is not full rank. In the latter case, Eq. (\ref{e15}) indicates that when eigenvalues $a_i$ of $\rho_{1}$ are all distinct, only one of the coefficients $b_i$, say
$b_l$, is zero and $a_{l}$ is the greatest eigenvalue of $\rho_{1}$, denoted as $a_{max}$.
Thus,
\begin{equation}\label{e19}
p_{opt}=\eta a_{max}d
\end{equation}
and $\pi_1=\beta_{l}|l\rangle\langle l|$ with $\beta_{l}$ an unknown positive constant and
\begin{equation}\label{e19-1}
\Pi_i=\lambda_{i}\beta_{l}U_{i}|l\rangle\langle l|U^{-1}_{i}.
\end{equation}
\par
Now, let there exist $m$ eigenvectors $i_1,\ldots, i_m$ of $\rho_1$ with the same eigenvalue $a_{max}$. Then, $p_{opt}$ is again given by (\ref{e19}) and by Eq. (\ref{e15}), these eigenvectors must correspond to the eigenvalue zero of
$\tau_1$. So, the operator $\pi_1$ can be written
as
\begin{equation}\label{e19-4}
\pi_1=\alpha_1|i_1\rangle \langle
i_1|+\ldots+\alpha_m|i_m\rangle \langle
i_m|,
\end{equation}
where $\alpha_i$'s are non-negative
numbers. By the same argument, we can get for the second set relations similar to Eqs. (\ref{e19}), (\ref{e19-1}) and (\ref{e19-4}). If this equality do not hold, all elements of one of sets $\{b_{i}\}^{n}_{i=1}$ and $\{b'_{i}\}^{n'}_{i=1}$ are nonzero.
\par
If $\tau_{1}$ ($\tau'_{1}$) be full rank, then it cannot be perpendicular to $\pi_{1}$ ($\pi'_{1}$) and by Eq. (\ref{e17}), $\lambda_{j}$'s ($\lambda'_{j}$'s) are all necessarily zero. When this is the case, operators of optimal POM corresponding to the first (second) set are all zero
and the unprimed (primed) states do not enter in the discrimination.\\
\subsection{The reducible case}
Let $U_i$'s and $U'_j$'s be generating sets of
reducible representations. Then, it is shown that the invariance of $\mathcal{M}$ under the operators
$U_i$'s and $U'_j$'s (see Eq. (\ref{e9-5})) requires that $\mathcal{M}$
is diagonal; i.e., $\mathcal{M}=diag(M_{1},\ldots,M_{d})$ (for a proof, see Appendix B of \cite{Jafarizadeh2}).\\
Here, the same technique as the irreducible case is applicable. However, we cannot give a general solution because the explicit form of $M$ differs  per case. In what follows, we illustrate the problem by considering some examples of qubit and $m$-qubit mixed states.\\
\section{Examples}
\textbf{I. MED between two sets of similarity transformed equiprobable qubit states}\\\\
Let us consider two different qubit states as
\begin{equation}
\rho_1=\frac{1}{2}(I+b\hat{n}^{(1)}.\vec{\sigma}),
\end{equation}
\begin{equation}\label{e20}
\rho'_1=\frac{1}{2}(I+b'\hat{n}'^{(1)}.\vec{\sigma}),
\end{equation}
where $\sigma_{i}$'s are the Pauli matrices and $\hat{n}^{(1)}$ and $\hat{n}'^{(1)}$ are unit vectors. Furthermore, let
$U_j$ and $U^{'}_j$ be arbitrary rotations about the
z-axis that rotate $\rho_1$ and $\rho'_1$ into $\rho_j$ and $\rho'_j$, respectively, such that
\begin{equation}\label{e20-1}
\rho_j=\frac{1}{2}(I+b\hat{n}^{(j)}.\vec{\sigma}),
\end{equation}
\begin{equation}\label{e20-2}
\rho'_j=\frac{1}{2}(I+b'\hat{n}'^{(j)}.\vec{\sigma}).
\end{equation}
The conjugate states corresponding to $\rho_{j}$ and $\rho^{'}_{j}$ have the form
\begin{equation}\label{e20-3}
\tau_j=\frac{1}{2}(I+c\hat{m}^{(j)}.\vec{\sigma}),
\end{equation}
\begin{equation}\label{e20-4}
\tau'_j=\frac{1}{2}(I+c'\hat{m}'^{(j)}.\vec{\sigma}).
\end{equation}
where $\hat{m}^{(j)}$ and $\hat{m}'^{(j)}$ are unit vectors. Now, from
the invariance of $\mathcal{M}$ under rotations about the $z$-axis,
it follows that $n>1$ or $n'>1$ and
\begin{equation}\label{e20-5}
\mathcal{M}=\alpha I_d+\beta\sigma_{z}.
\end{equation}
First, we assume that $n\geq2$ and $n'\geq2$. Hence, we have $p_{opt}>\eta,\eta'$ and by Eq. (\ref{e9-3}), $\lambda_{j}Tr(\tau_{j}\pi_{j})=\lambda'_{j}Tr(\tau'_{j}\pi'_{j})=0$. This consideration together with resolution of the identity imply
that at least one of $\tau_1$ and $\tau^{'}_1$ is not full rank. Without loss of generality, let us assume that $\tau_1$ is not
full rank such that its minimum eigenvalue $\frac{1-c}{2}$ is zero which yields
$c=1$.
To satisfy Eq. (\ref{e9-3}), the optimal POM elements must have the following expression
\begin{equation}\label{e26}
\Pi_j=\lambda_{j}U_{j}\pi_1U^{-1}_{j}=\lambda_{j}(I-m^{(j)}_{x}\sigma_{x}-m^{(j)}_{y}\sigma_{y}-m_{z}\sigma_{z}),\quad\lambda_{j}\geq0,
\end{equation}
\begin{equation}\label{e27}
\Pi'_j=\lambda'_{j}U_{j}\pi'_1U^{-1}_{j}=\lambda'_{j}(I-m'^{(j)}_{x}\sigma_{x}-m'^{(j)}_{y}\sigma_{y}-m'_{z}\sigma_{z}),\quad\lambda'_{j}\geq0.
\end{equation}
By substituting the expressions of $\Pi_j$ and $\Pi'_j$ from Eqs. (\ref{e26}) and (\ref{e27}) into Eq. (\ref{e9}),
it is easy to see that the following restrictions are imposed on $\lambda_{j}$ and $\lambda^{'}_{j}$
\begin{equation}\label{e30-1}
\sum^{n}_{j=1}\lambda_{j}+\sum^{n'}_{j=1}\lambda'_{j}=1,
\end{equation}
\begin{equation}\label{e30-2}
m_{z}\sum^{n}_{j=1}\lambda_{j}+m'_{z}\sum^{n'}_{j=1}\lambda'_{j}=0,
\end{equation}
\begin{equation}\label{e30-3}
\sum^{n}_{j=1}\lambda_{j}(m^{(j)}_{x}\textbf{i}+m^{(j)}_{y}\textbf{j})+\sum^{n'}_{j=1}\lambda'_{j}({m'}^{(j)}_{x}\textbf{i}+{m'}^{(j)}_{y}\textbf{j})=0.
\end{equation}
It is clear from Eqs. (\ref{e30-1}) and (\ref{e30-2}) that the signs of $m_{z}$ and $m'_{z}$ are opposite unless $m_{z}=m'_{z}=0$. Furthermore, Eqs. (\ref{e30-2}) and (\ref{e30-3}) imply that the points representing the optimal measurement operators do not share the same hemisphere of the
Bloch sphere.
\par
In what follows, it is convenient
to obtain relations between components of Bloch vectors of the states and whose conjugate states.
To this end, for the case $n>1$ and $n'>1$, we combine Eqs. (\ref{e9-4}) and (\ref{e20-1})-(\ref{e20-5}) and get
\begin{equation}\label{e30-4}
m^{(j)}_{x}=-\frac{\eta b}{p_{opt}-\eta}n^{(j)}_{x},\quad m^{(j)}_{y}=-\frac{\eta b}{p_{opt}-\eta}n^{(j)}_{y},\quad m_{z}=\frac{2\beta-\eta bn_{z}}{p_{opt}-\eta},
\end{equation}
\begin{equation}\label{e30-5}
c'm'^{(j)}_{x}=-\frac{\eta'b'}{p_{opt}-\eta'}n'^{(j)}_{x},\quad c'm'^{(j)}_{y}=-\frac{\eta'b'}{p_{opt}-\eta'}n'^{(j)}_{y},\quad c'm'_{z}=\frac{2\beta-\eta'b'n'_{z}}{p_{opt}-\eta'}.
\end{equation}
We discuss various situations based on whether $m_{z}=0$ ($m'_{z}=0$) or $m_{z}\neq0$ ($m'_{z}\neq0$). As we will see, the optimal measurement operators of one of the sets may become zero.\\\\
\textbf{\emph{1. $m_{z}\neq0$ and $m'_{z}\neq0$}}\\\\
Let $m_{z}\neq0$ and $m'_{z}\neq0$. Then, as mentioned above, the signs of $m_{z}$ and $m'_{z}$ must be opposite. Let us, without loss of generality, assume that $m_{z}>0$ and $m'_{z}<0$ which, by Eqs. (\ref{e30-4}) and (\ref{e30-5}), is equivalent to the condition
\begin{equation}\label{e30-7}
\frac{\eta bn_{z}}{2}<\beta<\frac{\eta'b'n'_{z}}{2}.
\end{equation}
In this case, the points of the Bloch sphere representing the optimal measurement operators of the first and second sets are in upper and lower hemispheres, respectively.
From this consideration and the fact that POM elements must hold a resolution of the identity, it follows that neither the optimal POM elements associated to the first nor to the second set are zero at a whole. Therefore, due to the orthogonality of any optimal measurement operator and its corresponding conjugate state, conjugate states of any set cannot be full rank and hence must be pure i.e. $c=c'=1$.
\par
To obtain $p_{opt}$, first we find $\beta$ by using Eqs. (\ref{e30-4}) and (\ref{e30-5}) in
\begin{equation}\label{e30-7-1}
\begin{array}{c}
(m^{(1)}_{x})^{2}+(m^{(1)}_{y})^{2}+(m_{z})^{2}=1, \\
(m'^{(1)}_{x})^{2}+(m'^{(1)}_{y})^{2}+(m'_{z})^{2}=1
\end{array}
\end{equation}
and subtracting the resulted equations. The result is
\begin{equation}\label{e30-8}
\beta=\frac{\eta^{2}(b^{2}-1)-{\eta'}^{2}({b'}^{2}-1)+2(\eta-\eta')p_{opt}}{4(\eta bn_{z}-\eta'b'n'_{z})},
\end{equation}
Then placing $\beta$ in the equation resulted from the first equation of (\ref{e30-7-1}), gives the desired result as
\begin{equation}\label{e30-8-1}
p_{opt}=\frac{-B\pm\sqrt{B^{2}-AC}}{2A},
\end{equation}
where
\begin{equation}
\begin{array}{c}
 \hspace{-5cm} A=(\eta-\eta')^{2}-(\eta bn_{z}-\eta'b'n'_{z})^{2}, \\
 \hspace{-4.2cm} B=[\eta^{2}(b^{2}-1)-{\eta'}^{2}({b'}^{2}-1)](\eta-\eta') \\
  \hspace{0.5cm} -2\eta bn_{z}(\eta-\eta')(\eta bn_{z}-\eta'b'n'_{z})+2\eta(\eta bn_{z}-\eta'b'n'_{z})^{2}, \\
   \hspace{-0.1cm}C=[\eta^{2}(b^{2}-1)-{\eta'}^{2}({b'}^{2}-1)]^{2}+4\eta^{2}(b^{2}-1)(\eta bn_{z}-\eta'b'n'_{z})^{2} \\
  -4\eta bn_{z}[\eta^{2}(b^{2}-1)-{\eta'}^{2}({b'}^{2}-1)](\eta bn_{z}-\eta'b'n'_{z}).
\end{array}
\end{equation}
Of course, from two roots of Eq. (\ref{e30-8-1}), we must take the biggest one as $p_{opt}$.\\\\
\textbf{\emph{2. $m_{z}=0$ and $m'_{z}=0$}}\\\\
When $m_{z}=m'_{z}=0$, from Eqs. (\ref{e30-4}) and (\ref{e30-5}) we have
\begin{equation}\label{e31}
\frac{\eta bn_{z}}{2}=\beta=\frac{\eta'b'n'_{z}}{2}.
\end{equation}
Here, the points representing optimal measurement operators are all on equator of the
Bloch Sphere. In this case, the success probability is given by
\begin{equation}\label{e32}
\mathcal{}p_{opt}=\eta(1+b\sqrt{1-n^{2}_{z}})
\end{equation}
and Eqs. (\ref{e30-4}) and (\ref{e30-5}) give $c'$ as
\begin{equation}\label{e32-1}
c'=\frac{\eta'b'}{p_{opt}-\eta'}\sqrt{1-n_{z}^{2}}.
\end{equation}
In the case of $c'<1$, all of $\lambda'_{j}$'s and hence primed measurement operators have to be zero while in the case of $c'=1$ it is not necessarily the case.\\\\
\textbf{\emph{3. $m_{z}=0$ and $m'_{z}\neq0$}}\\\\
In this case, Eqs. (\ref{e30-4}) and (\ref{e30-5}) lead to
\begin{equation}\label{e33}
\frac{\eta bn_{z}}{2}=\beta\neq\frac{\eta'b'n'_{z}}{2},
\end{equation}
For $n>1$, when this occurs, the success probability becomes as
\begin{equation}\label{e34}
p_{opt}=\eta(1+b\sqrt{1-n^{2}_{z}})
\end{equation}
and
\begin{equation}
c'=\frac{\sqrt{\eta^{2}b^{2}n^{2}_{z}+\eta'^{2}b'^{2}-2\eta\eta'bb'n_{z}n'_{z}}}{p_{opt}-\eta'}.
\end{equation}
Whether primed conjugate states are mixed, $c'<1$, or pure, $c'=1$,
the coefficients $\lambda'_{j}$'s and hence $\Pi'_{j}$'s should all be zero, since otherwise the measurement operators cannot satisfy a resolution of the identity. Here, the optimal measurement is the one which optimally distinguishes the states $\rho_{1},\ldots,\rho_{n}$ only.
Furthermore, Eq. (\ref{e30-3}) by using Eq. (\ref{e30-4}) imposes the constraint
\begin{equation}\label{e36}
\sum_{j=1}^{n}\lambda_{j}(n^{(j)}_{x}\textbf{i}+n^{(j)}_{y}\textbf{j})=0,
\end{equation}
which means that sates of the first set do not all place on the same half of the equator of the
Bloch sphere.
\par
In the case $n=1$, $n^{'}>1$, in order to satisfy the conditions (\ref{e9}) and (\ref{e9-3}),  we have $\lambda'_{j}=0$ for all $j$, and
\begin{equation}\label{e35}
p_{opt}=\eta,\quad\Pi_{1}=\mathrm{I}.
\end{equation}
\textbf{\emph{4. $m_{z}\neq0$ and $m'_{z}=0$}}\\\\
In this case, we have $c=c'=1$ and
\begin{equation}\label{e37}
\frac{\eta bn_{z}}{2}\neq\beta=\frac{\eta'b'n'_{z}}{2},
\end{equation}
Also, for $n'>1$, we obtain
\begin{equation}\label{e38}
p_{opt}=\eta'(1+b'\sqrt{1-n'^{2}_{z}}).
\end{equation}
\begin{equation}\label{e36}
\sum_{j=1}^{n'}\lambda'_{j}(n'^{(j)}_{x}\textbf{i}+n'^{(j)}_{y}\textbf{j})=0,
\end{equation}
and for $n>1$ and $n^{'}=1$, we have
\begin{equation}\label{e39}
p_{opt}=\eta',\quad\Pi'_{1}=\mathrm{I}.
\end{equation}
For the case $c\leq1$ and $c'=1$, the roles of parameters for the first and second set of states are reversed and the results are the same as the case $c=1$ and $c'\leq1$ only with unprimed and primed parameters exchanged.\\\\
\textbf{II. Some special cases}\\\\
In this section, we derive analytical expressions for the maximal success probability and optimal detection operators in some instances which extend and confirm the results obtained in Refs. \cite{Andersson1}-\cite{Jezek1}. Hereafter, we use the notations
\begin{equation}
\vec{\Lambda}^{(n)}=\sum^{n}_{j=1}\lambda_{j}(n^{(j)}_{x}\textbf{i}+n^{(j)}_{y}\textbf{j}),\quad\vec{\Lambda}'^{(n')}=\sum^{n'}_{j=1}\lambda'_{j}(n'^{(j)}_{x}\textbf{i}+n'^{(j)}_{y}\textbf{j}).
\end{equation}
\textbf{Case 1.}\\\\
Consider $n+1$ different pure states with the Bloch vectors
\begin{equation}\label{e39-1}
\hat{n}_{j}=(n^{(j)}_{x},n^{(j)}_{y},n_{z}),\quad\hat{n}'_{1}=(0,0,1),\quad j=1,\ldots,n.
\end{equation}
From Eqs. (\ref{e30-3}) and (\ref{e30-5}) we see that $m'^{(1)}_{x}=m'^{(1)}_{y}=0$ and $\vec{\Lambda}^{(n)}=0$.
Some simple algebra, gives
\begin{equation}\label{e39-2}
p_{opt}=\left\{
\begin{array}{ll}
\frac{2\eta'(\eta n_{z}-\eta')}{\eta(1+n_{z})-2\eta'} & $if$ \quad\eta n_{z}<\frac{2\eta'(\eta-\eta')}{\eta(1+n_{z})-2\eta'}<\eta'\quad $or$ \quad\eta n_{z}>\frac{2\eta'(\eta-\eta')}{\eta(1+n_{z})-2\eta'}>\eta',\\
&  \quad\quad $and$ \quad\quad\quad\frac{\eta}{p_{opt}-\eta}\vec{\Lambda}^{(n)}+\frac{\eta'}{p_{opt}-\eta'}\lambda'_{1}\textbf{j}=0;\\
\eta(1+\sqrt{1-n^{2}_{z}}) & $if$ \quad\eta n_{z}=\eta'\quad $or$ \quad\eta\leq\frac{n_{z}-1-\sqrt{1-n^{2}_{z}}}{n^{2}_{z}+\eta n_{z}-1-n-(1+n)\sqrt{1-n^{2}_{z}}},\quad\vec{\Lambda}^{(n)}=0.
\end{array}
\right.
\end{equation}
These results are reached by the following
sets of detection operators, respectively
\begin{equation}
\begin{array}{c}
  \Pi_j=\lambda_{j}[I+\frac{\eta}{p_{opt}-\eta}(n^{(j)}_{x}\sigma_{x}+n^{(j)}_{y}\sigma_{y})-\frac{(\eta-\eta')p_{opt}-\eta n_{z}(\eta n_{z}-\eta')}{(\eta n_{z}-\eta')(p_{opt}-\eta)}\sigma_{z}], \\
\hspace{-4.8cm}  \Pi'_1=\lambda'_{1}[I-\frac{(\eta-\eta')p_{opt}-\eta'(\eta n_{z}-\eta')}{(\eta n_{z}-\eta')(p_{opt}-\eta')}\sigma_{z}],
\end{array}
\end{equation}
and
\begin{equation}
\begin{array}{c}
\Pi_j=\lambda_{j}(I+\frac{n^{(j)}_{x}}{\sqrt{1-n^{2}_{z}}}\sigma_{x}+\frac{n^{(j)}_{y}}{\sqrt{1-n^{2}_{z}}}\sigma_{y}),\\
\hspace{-4.8cm}  \Pi'_1=0.
\end{array}
\end{equation}
\textbf{Case 2.}\\\\
Consider $n+1$ different pure states with the Bloch vectors
\begin{equation}\label{e39-3}
\hat{n}_{j}=(n^{(j)}_{x},n^{(j)}_{y},n_{z}),\quad\hat{n}^{'}_{1}=(0,1,0),\quad j=1,\ldots,n.
\end{equation}
Here, we obtain
\begin{equation}\label{e40-1}
p_{opt}=\left\{
  \begin{array}{ll}
    \frac{2\eta^{2}\eta^{'}n^{2}_{z}}{\eta^{2}n^{2}_{z}-(\eta-\eta^{'})^{2}} & $if$\quad0<\frac{(\eta-\eta')\eta'}{(\eta n_{z})^{2}-(\eta-\eta^{'})^{2}}<\frac{1}{2},\quad  \frac{\eta}{p_{opt}-\eta}\vec{\Lambda}^{(n)}+\frac{\eta'}{p_{opt}-\eta'}\lambda'_{1}\textbf{j}=0;\\
2\eta & $if$\quad n_{z}=0,\quad\eta\geq\frac{1}{1+n},\quad\vec{\Lambda}^{(n)}+\lambda'_{1}\textbf{j}=0;\\
    \eta(1+\sqrt{1-n^{2}_{z}}) & $if$\quad n_{z}\neq0,\quad\eta\geq\frac{1}{n+\sqrt{1-n^{2}_{z}}},\quad \vec{\Lambda}^{(n)}=0;\\
    \eta & $if$\quad n_{z}\neq0,\quad\eta=\frac{1}{1+n};
  \end{array}
\right.
\end{equation}
where are reached by the following
set of detection operators, respectively
\begin{equation}
  \begin{array}{ll}
 \Pi_{j}=\lambda_{j}[I+\frac{\eta}{p_{opt}-\eta}(n^{(j)}_{x}\sigma_{x}+n^{(j)}_{y}\sigma_{y})-\frac{(\eta-\eta')p_{opt}-\eta^{2}n^{2}_{z}}{\eta n_{z}(p_{opt}-\eta)}\sigma_{z}], \\
    \Pi'_{1}= \lambda'_{1}[I+\frac{\eta'}{p_{opt}-\eta'}\sigma_{y}-\frac{(\eta-\eta')p_{opt}}{\eta n_{z}(p_{opt}-\eta')}\sigma_{z}];
  \end{array}
\end{equation}
\begin{equation}
  \begin{array}{ll}
 \Pi_{j}=\lambda_{j}(I+n^{(j)}_{x}\sigma_{x}+n^{(j)}_{y}\sigma_{y}),\\
    \Pi'_{1}= \lambda'_{1}(I+\sigma_{y});
  \end{array}
\end{equation}
\begin{equation}
  \begin{array}{ll}
 \Pi_{j}=\lambda_{j}[I+\frac{1}{\sqrt{1-n^{2}_{z}}}(n^{(j)}_{x}\sigma_{x}+n^{(j)}_{y}\sigma_{y})],\\
    \Pi'_{1}= 0;
  \end{array}
\end{equation}
and
\begin{equation}
  \begin{array}{ll}
 \Pi_{j}=0,\\
    \Pi'_{1}=I.
  \end{array}
\end{equation}
\textbf{Case 3.}\\\\
As a final case, let us consider two set such that each one containing two pure states with Bloch vectors as follows
\begin{equation}\label{e41}
\hat{n}_{j}=(n^{(j)}_{x},n^{(j)}_{y},n_{z}),\quad\hat{n}^{'}_{j}=({n^{'}}^{(j)}_{x},{n^{'}}^{(j)}_{y},0),\quad j=1,2.
\end{equation}
By referring to the general case discussed above, the optimal success probability and measurement have given by
\begin{equation}\label{e44}
p_{opt}=\left\{
  \begin{array}{ll}
    \frac{2\eta^{2}\eta^{'}n^{2}_{z}}{\eta^{2}n^{2}_{z}-(\eta-\eta^{'})^{2}} & $if$\quad0<\frac{(\eta-\eta')\eta'}{(\eta n_{z})^{2}-(\eta-\eta^{'})^{2}}<\frac{1}{2},\quad\frac{\eta}{p_{opt}-\eta}\vec{\Lambda}^{(2)}+\frac{\eta'}{p_{opt}-\eta'}\vec{\Lambda}'^{(2)}=0;\\
    2\eta & $if$\quad n_{z}=0,\quad\eta\geq\frac{1}{4},\quad\vec{\Lambda}^{(2)}+
    \vec{\Lambda}'^{(2)}=0;\\
\eta(1+\sqrt{1-n^{2}_{z}}) & $if$\quad\ n_{z}\neq0,\quad\eta\geq\frac{1}{2+\sqrt{1-n^{2}_{z}}},\quad \vec{\Lambda}^{(2)}=0;\\
2\eta' & $if$\quad n_{z}=0,\quad\eta\leq\frac{1}{4},\quad\vec{\Lambda}^{(2)}+\vec{\Lambda}'^{(2)}=0;\\
2\eta' & $if$\quad n_{z}\neq0,\quad\eta\leq\frac{1}{4},\quad\vec{\Lambda}'^{(2)}=0;
\end{array}
\right.
\end{equation}
and the associated measurement operators are respectively given by\\
\begin{equation}
  \begin{array}{ll}
 \Pi_{j}=\lambda_{j}[I+\frac{\eta}{p_{opt}-\eta}(n^{(j)}_{x}\sigma_{x}+n^{(j)}_{y}\sigma_{y})-\frac{(\eta-\eta')p_{opt}-\eta^{2}n^{2}_{z}}{\eta n_{z}(p_{opt}-\eta)}\sigma_{z}],\\
     \Pi'_{j}=\lambda'_{j}[I+\frac{\eta'}{p_{opt}-\eta'}(n'^{(j)}_{x}\sigma_{x}+n'^{(j)}_{y}\sigma_{y})-\frac{(\eta-\eta')p_{opt}}{\eta n_{z}(p_{opt}-\eta')}\sigma_{z}];
  \end{array}
\end{equation}
\begin{equation}
  \begin{array}{ll}
 \Pi_{j}=\lambda_{j}[I+\frac{1}{2}(n^{(j)}_{x}\sigma_{x}+n^{(j)}_{y}\sigma_{y})],\\
     \Pi'_{j}=\lambda'_{j}(I+n'^{(j)}_{x}\sigma_{x}+n'^{(j)}_{y}\sigma_{y});
  \end{array}
\end{equation}
\begin{equation}
  \begin{array}{ll}
 \Pi_{j}=\lambda_{j}[I+\frac{1}{\sqrt{1-n^{2}_{z}}}(n^{(j)}_{x}\sigma_{x}+n^{(j)}_{y}\sigma_{y})],\\
     \Pi'_{j}=0;
  \end{array}
\end{equation}
\begin{equation}
  \begin{array}{ll}
 \Pi_{j}=\lambda_{j}(I+n^{(j)}_{x}\sigma_{x}+n^{(j)}_{y}\sigma_{y}),\\
    \Pi'_{j}= \lambda'_{j}(I+n'^{(j)}_{x}\sigma_{x}+n'^{(j)}_{y}\sigma_{y});
  \end{array}
\end{equation}
and
\begin{equation}
  \begin{array}{ll}
 \Pi_{j}=0,\\
     \Pi'_{j}=\lambda'_{j}(I+n'^{(j)}_{x}\sigma_{x}+n'^{(j)}_{y}\sigma_{y}).
  \end{array}
\end{equation}
\textbf{III. MED between two sets of similarity transformed $m$-qubit states in generalized Bloch sphere}\\\\
We consider particular $m$-qubit states in $d=2^m$
dimensional Hilbert space which possess properties similar to qubit
density matrices represented in Bloch sphere, and so we call them
generalized Bloch sphere states. Then the decomposition of these density matrices into a Bloch vector has,
in general, the following form:
\begin{equation}\label{e45}
\rho=\frac{1}{2^{m}}(I+a\sum_{i=1}^{2m+1}n_i\gamma_i)=\frac{1}{2^{m}}(I+a\hat{n}.\vec{\gamma}),
\end{equation}
where $\gamma_i$ for $i=1,2,\ldots, 2m+1$, known as Dirac matrices, are generators of special
orthogonal group $SO(2m+1)$, and represented as traceless Hermitian
matrices in a $2^m$-dimensional Hilbert space. That is, $\gamma_i$ are
maximally anticommuting set
which satisfy
\begin{equation}\label{e46}
\{\gamma_i,\gamma_j\}=2\delta_{ij}I_d.
\end{equation}
For a brief review about Dirac matrices and an explicit construction
of $\gamma_i$s, we refer the reader to \cite{Weinberg1} or the Appendix $A$
of \cite{Jafarizadeh3}. From the properties (\ref{e46}), it is easy to see that $(\hat{n}.\vec{\gamma})^2=I$ and so the
eigenvalues of $\rho$ are $\frac{1\pm a}{2^m}$. Therefore, the spectral decomposed form of
the density matrix $\rho$ is
\begin{equation}\label{e47}
\rho=\frac{1+a}{2^{m}}(\frac{I+\hat{n}.\vec{\gamma}}{2})+\frac{1-a}{2^{m}}(\frac{I-\hat{n}.\vec{\gamma}}{2})
\end{equation}
where $\frac{I+\hat{n}.\vec{\gamma}}{2}$ and
$\frac{I-\hat{n}.\vec{\gamma}}{2}$ are projection
operators (idempotents) to the degenerate eigenspaces corresponding to
the eigenvalues $\frac{1+a}{2^{m}}$ and $\frac{1-a}{2^{m}}$, respectively. Here, $U_i=e^{{i\sum_{_{i< j}}\theta_{ij}\gamma_i\gamma_j}}$  are spinor representations of the group $SO(2m+1)$. It should be noticed that $U_i$'s can also be
chosen as spinor representation of any subgroup of $SO(2m+1)$ of the
same rank $m$ (maximal subgroup) such as $SO(i_1)\otimes
SO(i_2)\otimes \ldots \otimes SO(i_{l})$, where
$(i_1,i_2,\ldots,i_{l})$ is an arbitrary partition of $2m+1$ to $l$
parties, i.e., $2m+1=i_1+i_2+\ldots+i_{l}$. We write $\rho_{j}$ as
\begin{equation}\label{e47-1}
\rho_{j}=\frac{1}{2^{m}}(I+b\sum_{i\in
S_V}n^{(j)}_i\gamma_i+b\sum_{i\in S_I}n_i\gamma_i),
\end{equation}
where $S_V$ and $S_I$ stand for index sets of Bloch vector components which are respectively variant and invariant under unitary similarity transformations by $U_j$'s.
All of the discussions about qubit states can be extended to this case.\\\\
\textbf{A. irreducible case}\\\\
When $U_i$'s and $U'_i$'s are irreducible representations, as in qubit case of subsection 2, $\mathcal{M}$ is proportional to identity. Also, by Schur's lemma, invariant part of the density operators and their associated conjugate states are just equal to $\frac{1}{2^{m}}I$. As in the qubit case, at least one of $c$ and $c'$ is equal to one and here we take $c=1$. Since the discussion of $c'<1$ is similar to the qubit one, we proceed with $c'=1$ only. Therefore, we have
\begin{equation}\label{e48}
\rho_j=\frac{1}{2^{m}}(I+b\sum_{i\in
S_V}n^{(j)}_i\gamma_i),
\end{equation}
\begin{equation}\label{e49}
\rho'_j=\frac{1}{2^{m}}(I+b'\sum_{i\in
S'_{V}}n'^{(j)}_i\gamma_i),
\end{equation}
\begin{equation}\label{e50}
\tau_j=\frac{1}{2^{m}}(I+\sum_{i\in S_V}m^{(j)}_i\gamma_i),
\end{equation}
\begin{equation}\label{e51}
\tau'_j=\frac{1}{2^{m}}(I+\sum_{i\in S'_V}m'^{(j)}_i\gamma_i).
\end{equation}
To satisfy Eq. (\ref{e9-3}), we must choose the optimal POM elements as
\begin{equation}\label{e52}
\Pi_j=\lambda_{j}U_{j}\pi_1U^{-1}_{j}=\lambda_{j}(I-\sum_{i\in S_V}m^{(j)}_i\gamma_i),
\end{equation}
\begin{equation}\label{e53}
\Pi'_j=\lambda'_{j}U_{j}\pi'_1U^{-1}_{j}=\lambda'_{j}(I-\sum_{i\in S'_V}m'^{(j)}_i\gamma_i).
\end{equation}
By using Eqs. (\ref{e9-4}) and (\ref{e48})-(\ref{e51}), the components of Bloch vectors associated with optimal measurement are found as
\begin{equation}\label{e54}
m^{(j)}_{i}=-\frac{\eta b}{p-\eta}n^{(j)}_{i},\quad\quad\quad\quad m'^{(j)}_{i}=-\frac{\eta'b'}{p-\eta'}n'^{(j)}_{i},\quad\quad\quad\quad\quad i\in S_{V}.
\end{equation}
Finally, from the fact that the vectors $\hat{m}^{(j)}$, $\hat{m}'^{(j)}$, $\hat{n}^{(j)}$ and $\hat{n}'^{(j)}$ have
unit lengths, the minimum error probability is found
\begin{equation}\label{e56}
p_{opt}=\eta(1+b)=\eta'(1+b').
\end{equation}
which is in agreement with the general result of irreducible case, i.e. Eq. (\ref{e19}), for $d=2^{m}$, $a_{max}=\frac{1+b}{2^{m}}$ and $a'_{max}=\frac{1+b'}{2^{m}}$.\\\\
\textbf{B. reducible case}\\\\
Let $U_j$'s and $U'_j$'s are reducible representations.
Since the operators $U_{j}$'s and $U'_{j}$'s commute with $\mathcal{M}$, the index set associated to the  maximal number of $\gamma_i$'s which remain invariant under similarity transformations produced by them is $S_{I}\cap S'_{I}$. Thus we have
\begin{equation}\label{e57}
\mathcal{M}=\alpha I_d+\sum_{i\in S_{I}\cap S'_{I}}\beta_{i}\gamma_{i}.
\end{equation}
As any optimal POM element is perpendicular to the corresponding conjugate state, hence,
for any $j$, we have
\begin{equation}\label{e61}
\Pi_j=\lambda_{j}U_{j}\pi_1U^{-1}_{j}=\lambda_{j}(I-\sum_{i\in S_V}m^{(j)}_i\gamma_i-\sum_{i\in S_I}m_i\gamma_i),
\end{equation}
\begin{equation}\label{e62}
\Pi'_j=\lambda'_{j}U_{j}\pi'_1U^{-1}_{j}=\lambda'_{j}(I-\sum_{i\in S'_V}m'^{(j)}_i\gamma_i-\sum_{i\in S'_I}m'_i\gamma_i).
\end{equation}
Placing these relations into the completeness relation (\ref{e9}), we obtain
\begin{equation}\label{e63}
\sum^{n}_{j=1}\lambda_{j}+\sum^{n'}_{j=1}\lambda'_{j}=1
\end{equation}
\begin{equation}\label{e64}
\sum^{n}_{j=1}\lambda_{j}m^{(j)}_{i}+\sum^{n'}_{j=1}\lambda'_{j}m'^{(j)}_{i}=0,\quad\quad i\in S_{V}\cap S'_{V}
\end{equation}
\begin{equation}\label{e65}
m_{i}\sum^{n}_{j=1}\lambda_{j}+m'_{i}\sum^{n'}_{j=1}\lambda'_{j}=0,\quad\quad i\in S_{I}\cap S'_{I}
\end{equation}
\begin{equation}\label{e66}
\sum^{n}_{j=1}\lambda_{j}m^{(j)}_{i}+m'_{i}\sum^{n'}_{j=1}\lambda'_{j}=0,\quad\quad i\in S_{V}-(S_{V}\cap S'_{V})
\end{equation}
\begin{equation}\label{e53-5}
m_{i}\sum^{n}_{j=1}\lambda_{j}+\sum^{n'}_{j=1}\lambda'_{j}m'^{(j)}_{i}=0,\quad\quad i\in S'_{V}-(S_{V}\cap S'_{V})
\end{equation}

By substituting $\mathcal{M}$ from Eq. (\ref{e57}) into Eq. (\ref{e9-4}), we conclude that
\begin{equation}\label{e57-1}
\begin{array}{c}
\hspace{-1.7cm}  m^{(j)}_{i}=-\frac{\eta b}{p_{opt}-\eta}n^{(j)}_{i},\quad\quad i\in S_{V}; \\\\
  m_{i}=-\frac{\eta b}{p_{opt}-\eta}n_{i},\quad\quad i\in S_{V}-(S_{V}\cap S'_{V}); \\\\
\hspace{-1.7cm}  m_{i}=\frac{2^{m}\beta_{i}-\eta bn_{i}}{p_{opt}-\eta},\quad\quad i\in S_{I}\cap S'_{I},
\end{array}
\end{equation}
and
\begin{equation}\label{e57-2}
\begin{array}{c}
\hspace{-1.7cm}  m'^{(j)}_{i}=-\frac{\eta'b'}{p_{opt}-\eta'}n'^{(j)}_{i},\quad\quad i\in S'_{V}; \\\\
  m'_{i}=-\frac{\eta'b'}{p_{opt}-\eta'}n'_{i},\quad\quad i\in S'_{V}-(S_{V}\cap S'_{V}); \\\\
\hspace{-1.7cm}  m'_{i}=\frac{2^{m}\beta_{i}-\eta'b'n'_{i}}{p_{opt}-\eta'},\quad\quad i\in S_{I}\cap S'_{I}.
\end{array}
\end{equation}\\
To derive $p_{opt}$ by solving Eqs. (\ref{e65}), (\ref{e57-1}) and (\ref{e57-2}), first we denote restrictions of $\hat{n}$ and $\hat{n}'$
to the subspace corresponding to the index set $S_{I}\cap S'_{I}$ by $\vec{n}_{0}$ and $\vec{n}'_{0}$ respectively. Next, to simplify the algebra, we choose a coordinate system in the subspace corresponding to the index set $S_{I}\cap S'_{I}$ such that the vector $\vec{n}_{0}$ is directed along an axis. In this coordinate system, let us denote by $n'_{0}$ the component of $\vec{n}'_{0}$ along $\vec{n}_{0}$ and by $n'_{1}$ its component along an axis in the plane of $\vec{n}_{0}$ and $\vec{n}'_{0}$ which is perpendicular to $\vec{n}_{0}$. By some tricky algebra which is discussed in detail in Appendix C, we find
\begin{equation}\label{pro1}
p_{opt}=\frac{-B\pm\sqrt{B^{2}-AC}}{2A},
\end{equation}
where
\begin{equation}\label{eABC}
\begin{array}{c}
\hspace{-7.1cm}  A=(\eta-\eta')^{2}-(\eta bn_{0}-\eta'b'n'_{0})^{2}-\eta'^{2}b'^{2}n'^{2}_{1} \\\\
\hspace{-5.7cm}  B=[\eta^{2}(1-b^{2})-\eta'^{2}(1-b'^{2})-\frac{2\eta\eta'^{2}bb'^{2}n_{0}n'^{2}_{1}}{\eta bn_{0}-\eta'b'n'_{0}}](\eta-\eta') \\\\
\hspace{-0.3cm}   -2\eta bn_{0}(\eta-\eta')(\eta bn_{0}-\eta'b'n'_{0}+\frac{\eta'^{2}b'^{2}n'^{2}_{1}}{\eta bn_{0}-\eta'b'n'_{0}})+2\eta[(\eta bn_{0}-\eta'b'n'_{0})^{2}+\eta'^{2}b'^{2}n'^{2}_{1}]\\\\
\hspace{-7cm}  C=[\eta^{2}(1-b^{2})-\eta'^{2}(1-b'^{2})
-\frac{\eta\eta'^{2}bb'^{2}n_{0}n'^{2}_{1}}{\eta bn_{0}-\eta'b'n'_{0}}]^{2}\\\\
\hspace{-3cm} +4\eta^{2}[b^{2}(1+\frac{\eta'^{2}b'^{2}n^{2}_{0}n'^{2}_{1}}{(\eta bn_{0}-\eta'b'n'_{0})^{2}})-1][(\eta bn_{0}-\eta'b'n'_{0})^{2}+\eta'^{2}b'^{2}n'^{2}_{1}] \\\\
\hspace{-0.5cm} -4\eta bn_{0}[\eta^{2}(1-b^{2})-\eta'^{2}(1-b'^{2})
-\frac{2\eta\eta'^{2}bb'^{2}n_{0}n'^{2}_{1}}{\eta bn_{0}-\eta'b'n'_{0}}](\eta bn_{0}-\eta'b'n'_{0}+\frac{\eta'^{2}b'^{2}n'^{2}_{1}}{\eta bn_{0}-\eta'b'n'_{0}}).
\end{array}
\end{equation}
It is easy to see that the statement of $p_{opt}$ is reduced to the statement of qubit case if we let $S_{I}=S'_{I}$ be one-dimensional and so $n'_{1}=0$, $n_{0}=n_{z}$ and $n'_{0}=n'_{z}$.
\section{Conclusion}
Using the necessary and sufficient
conditions for minimum-error discrimination in the equality form which is equivalent to Helstrom family of ensembles, we investigated minimum-error discrimination among two sets of different equiprobable quantum states where each set
generated from a density operator by unitary similarity transformations.
In the case that the unitary operators involved in the similarity transformations are generating of irreducible representations of two subgroups of $U(d)$, we precisely derived the maximum success probability
and the optimal measurement. However, for the case that the
representations are reducible, we did not solve optimality conditions in general and instead we
illustrated the method by applying it to solve optimality
conditions for two set of equiprobable qubit states together with some particular cases and two set of equiprobable m-qubit states. The presented particular cases were extended forms of some examples given in the literature and our results confirm theirs.
\newpage
 \vspace{1cm}\setcounter{section}{0}
 \setcounter{equation}{0}
 \renewcommand{\theequation}{A-\roman{equation}}
{\Large{Appendix A}}\\\\
Here, we want to calculate maximal success probability given in Eq. (\ref{pro1}). To this aim, first we express Eq. (\ref{e9-4}) in terms of the Bloch vectors components as
\begin{equation}\label{A1}
\eta bn_{i}-\eta'b'n'_{i}=(p_{opt}-\eta')m'_{i}-(p_{opt}-\eta)m_{i}
\end{equation}
\begin{equation}\label{A2}
2^{m}\beta_{i}=\eta bn_{i}+(p_{opt}-\eta)m_{i}
\end{equation}
for any $i\in S_{I}\cap S'_{I}$. From Eqs. (\ref{A1}) and (\ref{e65}), it is seen that for any $i\in S_{I}\cap S'_{I}$, the signs of $\eta bn_{i}-\eta'b'n'_{i}$ and $m_{i}$ are opposite. In other words, if restrictions of $\hat{n}$, $\hat{n}'$, $\hat{m}$ and $\hat{m}'$
to the subspace corresponding to the index set $S_{I}\cap S'_{I}$ are denoted by $\vec{n}_{0}$, $\vec{n}'_{0}$, $\vec{m}_{0}$ and $\vec{m}'_{0}$ respectively, then the vectors $\eta b\vec{n}_{0}-\eta'b'\vec{n}'_{0}$ and $\vec{m}_{0}$ point in the opposite directions and we can write
\begin{equation}\label{A2-1}
\vec{m}_{0}=-\sqrt{\sum_{i\in S_{I}\cap S'_{I}}(m_{i})^{2}}\frac{\eta b\vec{n}_{0}-\eta'b'\vec{n}'_{0}}{|\eta b\vec{n}_{0}-\eta'b'\vec{n}'_{0}|}.
\end{equation}
Therefore, Eq. (\ref{A2}) takes the vectorial form
\begin{equation}\label{A3}
2^{m}\vec{\beta}=\eta b\vec{n}_{0}-(p_{opt}-\eta)\sqrt{\sum_{i\in S_{I}\cap S'_{I}}(m_{i})^{2}}\frac{\eta b\vec{n}_{0}-\eta'b'\vec{n}'_{0}}{|\eta b\vec{n}_{0}-\eta'b'\vec{n}'_{0}|}
\end{equation}
Eqs. (\ref{e65}), (\ref{A2-1}) and (\ref{A3}) show that the vectors $\vec{n}_{0}$, $\vec{n}'_{0}$, $\vec{\beta}$, $\vec{m}_{0}$ and $\vec{m}'_{0}$ are coplanar. Next, to simplify the algebra, we choose an orthogonal coordinate system in the subspace corresponding to the index set $S_{I}\cap S'_{I}$ such the plane of $\vec{n}_{0}$ and $\vec{n}'_{0}$ coincide with the plane defined by an arbitrary pair of coordinate axes and $\vec{n}_{0}$ points to the positive
direction of one axis of the pair.
Let us denote by $n'_{0}$ and $\beta_{0}$ the components of $\vec{n}'_{0}$ and $\vec{\beta}$ along an axis of the pair lying in the direction of $\vec{n}_{0}$ and by $n'_{1}$ and $\beta_{1}$ their components along another axis, respectively.
In the considered frame, Eq. (\ref{e65}) and the third relations of Eqs. (\ref{e57-1}) and (\ref{e57-2}) are written as
\begin{equation}\label{A4}
\begin{array}{c}
  \mu m_{0}+(1-\mu)m'_{0}=0 \\
  \mu m_{1}+(1-\mu)m'_{1}=0
\end{array}
\end{equation}
\begin{equation}\label{A5}
\begin{array}{c}
  m_{0}=\frac{2^{m}\beta_{0}-\eta bn_{0}}{p_{opt}-\eta}, \\
  m_{1}=\frac{2^{m}\beta_{1}}{p_{opt}-\eta},\\
  m'_{0}=\frac{2^{m}\beta_{0}-\eta'b'n'_{0}}{p_{opt}-\eta'},\\
  m'_{1}=\frac{2^{m}\beta_{1}-\eta'b'n'_{1}}{p_{opt}-\eta'},
\end{array}
\end{equation}
where we have introduced $\mu=\sum^{n}_{j=1}\lambda_{j}$ and $m_{0}$ ($m'_{0}$) and $m_{1}$ ($m'_{1}$) are components of $\vec{m}_{0}$ ($\vec{m}'_{0}$). When we square both sides of the first two relations of Eqs. (\ref{e57-1}) and (\ref{e57-2}), then sum up over $i$ and use the unity of $\hat{n}$, $\hat{n}'$, $\hat{m}$ and $\hat{m}'$, we obtain
\begin{equation}\label{A6}
\begin{array}{c}
  1-m^{2}_{0}-m^{2}_{1}=\frac{\eta^{2}b^{2}}{(p_{opt}-\eta)^{2}}(1-n^{2}_{0}) \\
  1-m'^{2}_{0}-m'^{2}_{1}=\frac{\eta'^{2}b'^{2}}{(p_{opt}-\eta')^{2}}(1-n'^{2}_{0}-n'^{2}_{1})
\end{array}
\end{equation}
Finally, by composing Eqs. (\ref{A4})-(\ref{A6}) we attain to Eq. (\ref{pro1}) for $p_{opt}$.

\end{document}